\documentclass{article} 
\usepackage{arxiv}
\usepackage[hidelinks]{hyperref}       
\usepackage{url}            
\usepackage{booktabs}       
\usepackage{amsfonts}       
\usepackage{graphicx}
\usepackage{subcaption}
\usepackage{booktabs}
\usepackage{pgfplots} \pgfplotsset{compat=1.9} \usepgfplotslibrary{fillbetween}
\usepackage{siunitx}
\usepackage[acronym]{glossaries}
\usepackage[numbers]{natbib}

\usetikzlibrary{calc}

\usepackage[normalem]{ulem}

\title{Data-Based In-Cylinder Pressure Model with Cyclic Variations for Combustion Control: A RCCI Engine Application}

\newif\ifuniqueAffiliation
\uniqueAffiliationtrue

\ifuniqueAffiliation 
\author{ \href{https://orcid.org/0000-0002-8343-8176}{\includegraphics[scale=0.06]{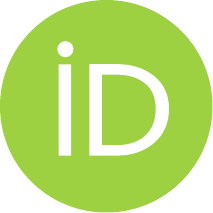}\hspace{1mm}Maarten Vlaswinkel} \\
	Department of Mechanical Engineering\\
	Eindhoven University of Technology\\
	Eindhoven, The Netherlands \\
	\texttt{m.g.vlaswinkel@tue.nl} \\
	\And
	\href{https://orcid.org/0000-0002-3550-4001}{\includegraphics[scale=0.06]{orcid.pdf}\hspace{1mm}Frank Willems} \\
	Department of Mechanical Engineering\\
	Eindhoven University of Technology\\
	Eindhoven, The Netherlands \\
    \texttt{f.p.t.willems@tue.nl} \\ \\
    Powertrains Department\\
    TNO Mobility \& Built Environment\\
    Helmond, The Netherlands
}
\else
\usepackage{authblk}

\newbox{\orcid}\sbox{\orcid}{\includegraphics[scale=0.06]{orcid.pdf}} 
\author[1]{%
	\href{https://orcid.org/0000-0000-0000-0000}{\usebox{\orcid}\hspace{1mm}David S.~Hippocampus\thanks{\texttt{hippo@cs.cranberry-lemon.edu}}}%
}
\author[1,2]{%
	\href{https://orcid.org/0000-0000-0000-0000}{\usebox{\orcid}\hspace{1mm}Elias D.~Striatum\thanks{\texttt{stariate@ee.mount-sheikh.edu}}}%
}
\affil[1]{Department of Computer Science, Cranberry-Lemon University, Pittsburgh, PA 15213}
\affil[2]{Department of Electrical Engineering, Mount-Sheikh University, Santa Narimana, Levand}
\fi

\newacronym{acr:ltc}{LTC}{Low Temperature Combustion}
\newacronym{acr:rcci}{RCCI}{Reactivity Controlled Compression Ignition}
\newacronym{acr:com}{COM}{Control-oriented Model}
\newacronym{acr:pc}{PC}{Principle Component}
\newacronym{acr:pcd}{PCD}{Principle Component Decomposition}
\newacronym{acr:pca}{PCA}{Principle Component Analysis}
\newacronym{acr:gpr}{GPR}{Gaussian Process Regression}
\newacronym{acr:cpbc}{CPBC}{Cylinder Pressure-Based Control}
\newacronym{acr:ivc}{IVC}{Intake Valve Close}
\newacronym{acr:pfi}{PFI}{Port Fuel Injection}
\newacronym{acr:di}{DI}{Direct Injection}
\newacronym{acr:egr}{EGR}{Exhaust Gas Recirculation}
\newacronym{acr:se}{SE}{Square Exponential}
\newacronym{acr:rq}{RQ}{Rational Quadratic}
\newacronym{acr:ard}{ARD}{Automatic Relevance Determination}
\newacronym{acr:mae}{MAE}{Mean Absolute Error}
\newacronym{acr:icc}{ICC}{In-Cylinder Conditions}

\newcommand{\trans}{\mathsf{T}}

\begin{document} 
\maketitle

\begin{abstract}
Cylinder pressure-based control is a key enabler for advanced pre-mixed combustion concepts. Besides guaranteeing robust and safe operation, it allows for cylinder pressure and heat release shaping. This requires fast control-oriented combustion models. 
Over the years, mean-value models have been proposed that can predict combustion measures (e.g.,  Gross Indicated Mean Effective Pressure ($\text{IMEP}_\text{g}$), or the crank angle where 50\% of the total heat is released (CA50)) or models that predict the full in-cylinder pressure. However, these models are not able to capture cycle-to-cycle variations. The inclusion of the cycle-to-cycle variations is important in the control design for combustion concepts, like Reactivity Controlled Compression Ignition, that can suffer from large cycle-to-cycle variations. In this study, the in-cylinder pressure and cycle-to-cycle variation are modelled using a data-based approach. The in-cylinder conditions and fuel settings are the inputs to the model. The model combines Principle Component Decomposition and Gaussian Process Regression. A detailed study is performed on the effects of the different hyperparameters and kernel choices. The approach is applicable to any combustion concept, but most valuable for advance combustion concepts with large cycle-to-cycle variation. The potential of the proposed approach is demonstrated for
an Reactivity Controlled Compression Ignition engine running on Diesel and E85. 
The prediction quality of the evaluated combustion measures has an overall accuracy of 13.5\% and 65.5\% in mean behaviour and standard deviation, respectively. The peak-pressure rise-rate is traditionally hard to predict, in the proposed model it has an accuracy of 22.7\% and 96.4\% in mean behaviour and standard deviation, respectively.
This Principle Component Decomposition-based approach is an important step towards in-cylinder pressure shaping. The use of Gaussian Process Regression provides important information on cycle-to-cycle variation and provides next-cycle controls information on safety and performance criteria.
\end{abstract}


\section{Introduction}
Concerns about global warming require a significant reduction in $\text{CO}_2$ emissions for on-road applications. 
This resulted in the interest of high efficient and low carbon propulsion methods in the transportation sector. This trend resulted in electrification for personal mobility, but the go-to technology for heavy-duty applications has not yet been decided. 
High efficiency and clean internal combustion engines together with sustainable fuels are expect to play a significant role in the future \cite{leach_scope_2020a,duartesouzaalvarengasantos_internal_2021,benajes_review_2024}. Advanced combustion concepts provide promising solutions to increase efficiency. Concepts like Homogeneous Charge Compression Ignition, Partial Premixed Combustion and \gls{acr:rcci} have been proposed \cite{dempsey_comparison_2014}. From these methods, \gls{acr:rcci} provides high efficiency and fuel flexibility as well as controllability. \gls{acr:rcci} uses a combination of a low and high reactive fuel during combustion \cite{reitz_review_2015}. By changing the ratio between low and high reactivity fuel and their injection timing, it is possible to optimise combustion phasing, duration an magnitude. However, continuous monitoring of the combustion process and regulating this ratio and timing is required to guarantee robust and safe operation \cite{paykani_reactivity_2021}.

A key concept for enabling these advanced combustion concepts is \gls{acr:cpbc}  \cite{willems_cylinder_2018}. \gls{acr:cpbc} is a next-cycle control method that uses measured in-cylinder pressure to regulate the actuators. Several \gls{acr:cpbc} methods have already been proposed in literature \cite{verhaegh_datadriven_2022}. The proposed methods use measures derived from the in-cylinder pressure, e.g., Gross Indicated Mean Effective Pressure, peak pressure, peak pressure rise-rate or the moment when 50\% of the heat is released. However, these measures have no direct relation to thermal efficiency. Direct in-cylinder pressure or heat release shaping is required to have direct control over the efficiency, load and guarantee safe operation.

The move to model-based \gls{acr:cpbc} requires a \gls{acr:com} of the in-cylinder pressure. These \glspl{acr:com} can help in improving controller design and calibration, and can be embedded in the controller. The model should give a relation between the in-cylinder mixture composition, intake manifold pressure and intake manifold temperature, and the resulting in-cylinder pressure. The computation time of this \gls{acr:com} should be below the duration of a combustion cycle to make sure a new control action has been determined before the start of the next combustion cycle. In the case of \gls{acr:rcci}, a description of the cycle-to-cycle variations should be present in the \gls{acr:com}.

A distinction can be made between two types of models: first-principle physics-based models and data-based models. Physics-based models use first-principle physical relations to capture the combustion behaviour. On the other side, purely data-based models use black box modelling methods where measurements are used to create a mapping from input to output.

To model important combustion measures (e.g., Gross Indicated Mean Effective Pressure ($\text{IMEP}_\text{g}$), or crank angle where 50\% of the total heat is released (CA50)) basic first-principle models have been proposed \cite{khodadadisadabadi_modeling_2016,guardiola_combustion_2018,raut_dynamic_2018,kakoee_modeling_2020}. These models provided a deterministic and dynamic view of the relation between actuation and combustion measures without determining the full in-cylinder pressure. To add new combustion measures these models should be extended. This can be time consuming and reduces the flexibility of these models during combustion control development.

To model the full in-cylinder pressure, more complex first-principle models have been proposed. These include the multi-zone model of \citeauthor{bekdemir_controloriented_2015} \cite{bekdemir_controloriented_2015} or the fluid dynamic model of \citeauthor{klos_investigation_2015} \cite{klos_investigation_2015}. The complexity of these models result in computation times that exceeds the combustion time and are therefore not directly suited as \gls{acr:com}. A reduction in computation time is achieved by using static, data-driven, deterministic regression models to capture the behaviour of important combustion measures.

On the other hand, data-based models have been developed. A \gls{acr:gpr} model to map in-cylinder conditions to combustion measures has been proposed by \citeauthor{xia_robust_2020} \cite{xia_robust_2020}. A state-space model identified using data to model combustion phasing and peak pressure rise-rate has been proposed by \citeauthor{basina_datadriven_2020} \cite{basina_datadriven_2020}. These models are made to only provide information on the modelled combustion measures. Therefore, the model has to be extended to include other measures.

Capturing the full in-cylinder pressure using data, \gls{acr:pcd} models have been proposed. These models consist of a weighted sum of principle components where the weights are modelled using regression methods. \citeauthor{pan_unsupervised_2019} \cite{pan_unsupervised_2019} use a deterministic neural network to capture the behaviour of the weights. On the other hand, \citeauthor{vlaswinkel_databased_2022a} \cite{vlaswinkel_databased_2022a} use \gls{acr:gpr} model to capture the behaviour of the weights. The later approach makes it possible to include cycle-to-cycle variation in the model. 

The use of the \gls{acr:pcd} of the in-cylinder pressure has already been proposed in several control and detection methods. \citeauthor{henningsson_virtual_2012} \cite{henningsson_virtual_2012} used this decomposition as input to a virtual emission sensor. They where able to predict the air-to-fuel ratio and $\text{NO}_x$ emissions quite accurately. \citeauthor{panzani_engine_2017} \cite{panzani_engine_2017} and \citeauthor{panzani_engine_2019a} \cite{panzani_engine_2019a} proposed this decomposition for knock detection and avoidance. They used the decomposition to derive a measure of closeness to engine knocking. \citeauthor{vlaswinkel_cylinder_2023} \cite{vlaswinkel_cylinder_2023} used this decomposition as an alternative method to maximise the thermal efficiency. They used the decomposition to derive a measure of closeness of a measured in-cylinder pressure to an idealised thermodynamic cycle.

In this study, we will extend the work of \citeauthor{vlaswinkel_databased_2022a} \cite{vlaswinkel_databased_2022a} by giving an extensive analysis on: 
1) the comparison of different kernels in the \gls{acr:gpr} approach with regards to prediction quality of important combustion measures; 
2) understanding the effects of modelling correlated process as uncorrelated Gaussian process; 
3) using a data set with a wide range of operating conditions to show the effectiveness of the model.
This work is organised as follows. In Section~\ref{sec:exp_setup} an overview is given of the experimental setup and the used data set. Section~\ref{sec:comb_model} describes the data-based combustion model including cycle-to-cycle variation. A detailed analysis of the effect on different hyperparameters is presented in Section~\ref{sec:hp_select}. The prediction quality of the combustion model is demonstrated and validated in Section~\ref{sec:validation}.

\section{Single Cylinder Engine} \label{sec:exp_setup}
In this section, we will give a description of the setup and the used data sets. A discussion is provided on the chosen inputs to the model and how these are determined.

\subsection{System Description}
\begin{figure}
    \centering
    \includegraphics{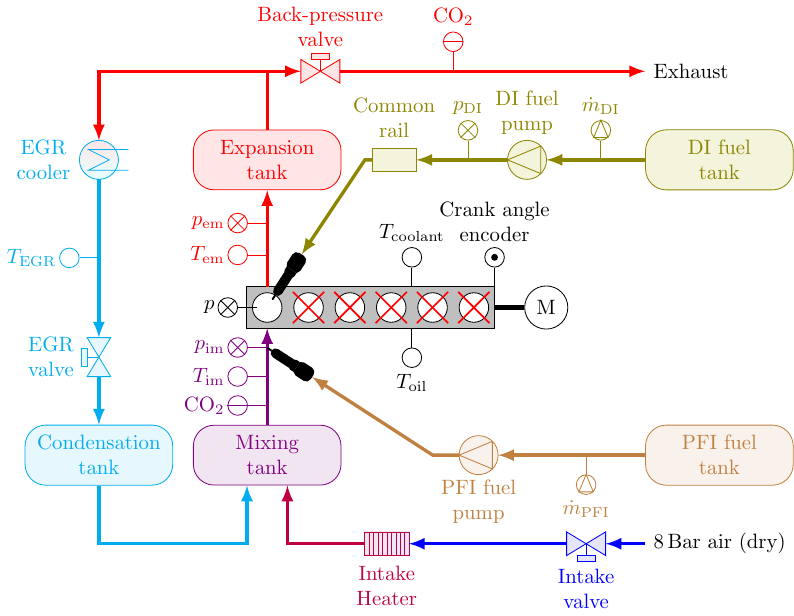}
    \caption{Schematic of the single cylinder PACCAR MX13 engine equipped with \acrlong{acr:egr} (\acrshort{acr:egr}), \acrlong{acr:di} (\acrshort{acr:di}) and \acrlong{acr:pfi} (\acrshort{acr:pfi}).} \label{fig:engine_setup}  
\end{figure}
\begin{table}[b]
	\centering
	\caption{Specifications of the engine setup} \label{tab:engineSpecs}
    \begin{tabular}{l l}
        \toprule
        Parameters & Value \\ \midrule
        Crank angle resolution & \SI{0.2}{\degree} \\
    	PFI fuel  & E85 \\	
		DI fuel  & Diesel (EN590)\\	
		Compression ratio  & 15.85 \\	
		Intake valve closure & \ang{-153}CA~aTDC   \\	
    	Exhaust valve opening & \ang{128}CA~aTDC  \\
    	Engine speed   & \SI{1200}{rpm}\\
        Oil temperature & \SI{90}{\degree C} \\
        Coolant temperature & \SI{87}{\degree C}
        \\ \bottomrule
    \end{tabular}
\end{table}
In this study, a modified PACCAR MX13 engine is used as shown in Figure~\ref{fig:engine_setup}. Cylinders 2 to 6 have been removed and only cylinder 1 is operational. To keep the engine running at a constant speed, the electric motor of the engine dynamometer provides the require torque. The focus is on \gls{acr:rcci} combustion with a single injection of diesel to auto-ignite the well-mixed charge of E85, air and recirculated exhaust gas. The injection of diesel does not ignite the mixture itself, but the ignition is caused by the increased temperature as a result of cylinder compression. Therefore, there is a clear temporal separation between the ignition of diesel and combustion. The \gls{acr:di} of diesel is handled by a Delphi DFI21 injector connected to a common rail. The E85 \gls{acr:pfi} is handled by a Bosch EV14 injector fitted into the intake channel set at \SI{5}{Bar}. Both the \gls{acr:di} and \gls{acr:pfi} fuel mass flows are measured using a Siemens Sitrans FC Mass 2100 Coriolis mass flow meter coupled with Mass 6000 signal converters. Boosted intake air is supplied at \SI{8}{Bar} and the pressure and temperature is regulated using a pressure regulator and an electric heater, respectively. The \gls{acr:egr} fraction is regulated by the \gls{acr:egr} and back-pressure butterfly valves. The \gls{acr:egr} flow is cooled down to approximately room temperature by a cooled stream of process water. The condensation tank collects the condensation form the \gls{acr:egr} flow and is drained regularly. The expansion and mixing tank dampen pressure fluctuations in the intake and exhaust manifold as a result of single cylinder operation. The in-cylinder pressure is sampled at \ang{0.2}~CA with a Kistler 6125C uncooled pressure transducer and amplified with a Kistler 5011B. A Leine Linde RSI 503 encoder provides crank angle information at a \SI{0.2}{\degree} interval. A Bronkhorst IN-FLOW F-106BI-AFD-02-V digital mass flow meter is used to measure the mass of the intake air flow. The pressure and temperatures located at different locations in the air-path are measured every combustion cycle using a Gems Sensors \& Controls 3500 Series pressure transmitter and Type-K thermocouples, respectively. The concentration of $\text{CO}_2$ in the intake and exhaust flows are measured using an Horiba MEXA 7100 DEGR system. 
Table~\ref{tab:engineSpecs} shows the specifications of the engine setup.

\subsection{Data Set for Model Training and Validation}
The model relates in-cylinder conditions, determined at intake valve closing, to a resulting in-cylinder pressure. These conditions consist of a range of parameters related to engine speed, cylinder wall temperature, and mixture composition, pressure and temperature. Since the engine is running at a single speed and at steady-state conditions the most relevant changes throughout the data set are a result of differences in mixture composition, pressure and temperature. These can be described using intake and fuelling conditions.
The chosen measurable parameters used to describe in-cylinder conditions are:
\begin{itemize}
	\item Total injected energy 
		\begin{equation*}
			Q_\text{total} = m_\text{PFI} \text{LHV}_\text{PFI} + m_\text{DI} \text{LHV}_\text{DI},
		\end{equation*}
        where $m_\text{PFI}$ and $m_\text{DI}$ are the injected masses of \gls{acr:pfi} and \gls{acr:di} fuels, and $\text{LHV}_\text{PFI}$ and $\text{LHV}_\text{DI}$ are the lower heating values of the  \gls{acr:pfi} and \gls{acr:di} fuels;
	\item Energy-based blend ratio 
		\begin{equation*}
			\text{BR} = \frac{m_\text{PFI} \text{LHV}_\text{PFI}}{Q_\text{total}};
		\end{equation*}
	\item Start-of-injection of the directly injected fuel $\text{SOI}_\text{DI}$;
	\item Pressure at the intake manifold $p_\text{im}$;
	\item Temperature at the intake manifold $T_\text{im}$; and 
	\item \gls{acr:egr} ratio 
		\begin{equation*}
			X_\text{EGR} = \frac{\text{CO}_{2,\text{in}}}{\text{CO}_{2,\text{out}}}
		\end{equation*}
		with $\text{CO}_{2,\text{in}}$ and $\text{CO}_{2,\text{out}}$ the concentration of $\text{CO}_2$ as a fraction of the volume flow at the intake and exhaust, respectively. 
\end{itemize}

\begin{figure}[!t]
    \includegraphics{ 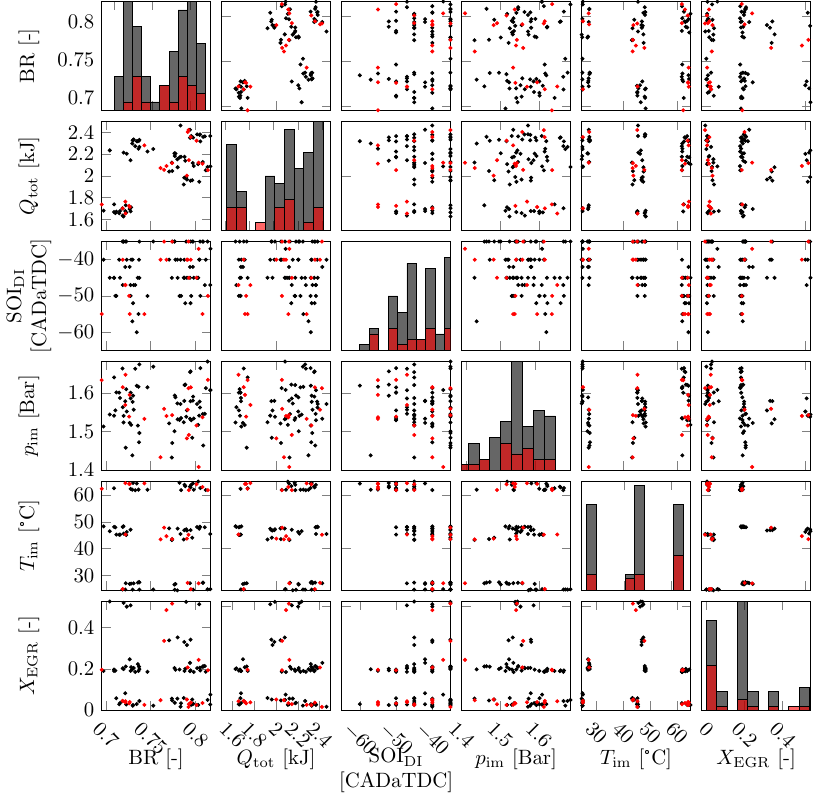 }
    \caption{Distribution of the in-cylinder conditions of the training (black) and validation data (red)} \label{fig:input_scatter}
\end{figure}
The variation in the in-cylinder conditions for the training data and validation data is shown in Figure~\ref{fig:input_scatter}. The diagonal shows the distribution of each measure for the in-cylinder conditions. The off-diagonal shows the joint distribution of the measures used for the in-cylinder conditions. The data set contains 95 different measurements consisting of $n_\text{cyc} = 50$ consecutive cycles each. Both small and large cycle-to-cycle variations, and non-firing behaviour are present within the data set. In this work, each cycle is used and no averaging over the $n_\text{cyc}$ in-cylinder conditions and in-cylinder pressure traces in a measurement is performed before analysis. The data set is randomly split into a training set of $n_\text{train} = 75$ measurements and a validation set of the remaining $n_\text{val} = 20$ measurements.


\section{Combustion Model} \label{sec:comb_model}
In this section, the data-based approach to model the in-cylinder pressure is introduced. It is based on the method presented in \citeauthor{vlaswinkel_databased_2022a} \cite{vlaswinkel_databased_2022a}. The approach combines \acrlong{acr:pcd} (\acrshort{acr:pcd}) and \acrlong{acr:gpr} (\acrshort{acr:gpr}). To describe the in-cylinder pressure during the compression and power stroke, the \gls{acr:pcd} is used to minimise the amount of information required by separating the influence of the in-cylinder conditions $s_\text{ICC}$ and the crank angle $\theta$ into two different mappings. \gls{acr:gpr} gives the possibility to model the in-cylinder pressure and cycle-to-cycle variation at different in-cylinder conditions.

\subsection{Principle Component Decomposition of the In-Cylinder Pressure} \label{sec:pca}
The in-cylinder pressure $p(\theta,\,s^*_\text{ICC})$ at crank angle $\theta \in \{\SI{-180}{\degree},\,\SI{-180}{\degree} + \Delta\text{CA},\, \dots,$ $\SI{180}{\degree} - \Delta\text{CA},\, \SI{180}{\degree}\}$ with $\Delta\text{CA}$ the crank angle resolution is decomposed as
\begin{equation} \label{eq:pcd_pressure}
    p(\theta,\,s^*_\text{ICC}) = p_\text{mot}(\theta,\,s^*_\text{ICC}) + w(s^*_\text{ICC})^\trans f(\theta),
\end{equation}
where $w(s^*_\text{ICC})$ is a vector of weights and $f(\theta)$ is the vector of principle components. In these vectors, the $i$th element is related to the $i$th \gls{acr:pc}. The in-cylinder condition $s^*_\text{ICC} \in \mathcal{S}^* \subset \mathcal{S}$ are in the set $\mathcal{S}^*$ containing all in-cylinder conditions present in the training set and the set $\mathcal{S}$ spanning the modelled operation domain.
It is assumed that the in-cylinder pressure during the intake stroke is equal to $p_\text{im}$.

The \glspl{acr:pc} are computed using the eigenvalue method. The $n_\text{train}\cdot n_\text{cyc}$ in-cylinder pressures $p(\theta,\,s^*_\text{ICC})$ contained in the training set are used. The vector $F_i$ is the $i$th unit eigenvector of the matrix $P P^\trans$ where $P \in \mathbb{R}^{n_\text{CA} \times n_\text{train}n_\text{cyc}}$ with $n_\text{CA}$ the number of crank angle values. The elements in matrix $P$ are defined as
\begin{equation}
    [P]_{ab} := p( \theta_a,\, s^*_{\text{ICC},b} ) -  p_\text{mot}(\theta_a,\,s^*_{\text{ICC},b}),
\end{equation}
such that the $a$th row of $P$ contains the values of the in-cylinder pressure at the $a$th crank angle for all $s_\text{ICC}^* \in \mathcal{S}^*$ and the $b$th column of $P$ contains the full in-cylinder pressure at all $\theta \in \{\SI{-180}{\degree},\,\SI{-180}{\degree} + \Delta\text{CA},\, \dots,\, \SI{180}{\degree} - \Delta\text{CA},\, \SI{180}{\degree}\}$ for the $b$th $s_\text{ICC}^*$.
The $i$th \gls{acr:pc} is defined as
\begin{equation} \label{eq:pcd_weight}
    f_i(\theta_a) = [F_i]_a.
\end{equation}
The weight related to the $i$th \gls{acr:pc} is given by
\begin{equation}
    w_i(s^*_\text{ICC}) = P(s^*_\text{ICC}) F_i,
\end{equation}
where $[P(s^*_\text{ICC})]_a = p(\theta_a,\,s^*_\text{ICC}) - p_\text{mot}(\theta_a,\,s^*_\text{ICC})$.
The training set generates a single set of \glspl{acr:pc}. These \glspl{acr:pc} are ordered by relevance, where $i=1$ is the most relevant \gls{acr:pc}. The determination of the \glspl{acr:pc} and the required amount of \glspl{acr:pc} will be done later in this study.

\subsection{Gaussian Process Regression to Capture Effects of In-Cylinder Conditions} \label{sec:gpr}
\gls{acr:gpr} is used to estimate the behaviour of $w(s_\text{ICC})$ over the full operation domain $\mathcal{S}$. To include cycle-to-cycle variations, $w(s_\text{ICC})$ is described by a stochastic process as
\begin{equation} \label{eq:gpr_w}
	w(s_\text{ICC}) := \mathcal{N}(\hat{w}(s_\text{ICC}),\,W(s_\text{ICC}))
\end{equation}
with mean $\hat{w}(s_\text{ICC}) := \mathbb{E}[w(s_\text{ICC})]$ and variance $W(s_\text{ICC}) := \mathbb{E}[ (w(s_\text{ICC}) - \hat{w}(s_\text{ICC}))(w(s_\text{ICC}) $\\$-\ \hat{w}(s_\text{ICC}))^\trans ]$. During this study, the correlation between output variable will be neglected (i.e., $,W(s_\text{ICC})$ is a diagonal matrix), since most literature on \gls{acr:gpr} assumes the output variables to be uncorrelated. This might effect the quality of the prediction of the cycle-to-cycle variation.

To improve the prediction accuracy and determination of the hyperparameters, normalised in-cylinder conditions $\bar{s}_\text{ICC}$ and weights $\bar{w}_i(s^*_\text{ICC})$ will be used. The in-cylinder condition scaling uses the mean $\bar{\mu}_{s^*_{\text{ICC}},j}$ and standard deviation $\bar{\sigma}_{s^*_{\text{ICC}},j}$ of the $j$th in-cylinder conditions variable over the full training set $S^*$ as
\begin{equation}
    \bar{s}_{\text{ICC},j} = \frac{ s_{\text{ICC},j} - \bar{\mu}_{s^*_{\text{ICC}},j} }{ \bar{\sigma}_{s^*_{\text{ICC}},j} }.
\end{equation}
The weight scaling uses the mean $\bar{\mu}_{w^*_{\text{ICC}},i}$ and standard deviation $\bar{\sigma}_{w^*_{\text{ICC}},i}$ of the $i$th in-cylinder conditions variable over the full training set $S^*$ as
\begin{equation}
    \bar{w}_i(s^*_\text{ICC}) = \frac{ w_i(s^*_\text{ICC}) - \bar{\mu}_{w_i} }{ \bar{\sigma}_{w_i} }.
\end{equation}
Following \cite{rasmussen_gaussian_2005}, the scaled expected value and scaled covariance matrix without correlation can be computed as:
 \begin{equation} \label{eq:mvgpr_exp}
    \hat{\bar{w}}_i(\bar{s}_\text{ICC}) = K( \bar{s}_\text{ICC},\,\bar{s}^*_\text{ICC},\,\phi ) \left(K(\bar{s}^*_\text{ICC},\,\bar{s}^*_\text{ICC},\,\phi) + \varphi_\text{n}I\right)^{-1} \bar{w}_i(\bar{s}^*_\text{ICC})
\end{equation}
and
\begin{equation} \label{eq:mvgpr_covar}
    \begin{aligned}
    \bar{W}_{ii}(\bar{s}_\text{ICC}) =\ &K( \bar{s}_\text{ICC},\, \bar{s}_\text{ICC},\,\phi ) \ - \\  &K( \bar{s}_\text{ICC},\,\bar{s}^*_\text{ICC},\,\phi ) \left(K(\bar{s}^*_\text{ICC},\,\bar{s}^*_\text{ICC},\,\phi) + \varphi_\text{n}I\right)^{-1} K^{\trans}( \bar{s}_\text{ICC} ,\, \bar{s}^*_\text{ICC},\,\phi),
    \end{aligned}
\end{equation}
where $K( \cdot,\,\cdot,\,\phi)$ is the kernel and $\phi$ and $\varphi_\text{n}$ are the kernel's hyperparameters. The selection of both elements will be discussed in the next section. 

To optimise the set of hyperparameters $\phi$  and $\varphi_\text{n}$ found in the kernels, the marginal log-likelihood is maximised for each \gls{acr:pc} separately. The marginal log-likelihood is often used in determining the hyperparameters in \gls{acr:gpr} and does not depend on the kernel type . It is given by
\begin{equation} \label{eq:mllo}
 \ln\left( \text{Prob}(\bar{w}_i\,|\,\bar{s}^*_\text{IVC},\,\phi) \right) = -\frac{1}{2} \bar{w}_i^\trans K_{\bar{s}^*_\text{IVC}}^{-1} \bar{w}_i - \frac{1}{2}\ln(\det(K_{\bar{s}^*_\text{IVC}})) - \frac{n_\text{exp}n_\text{cyc}}{2}\ln(2\pi),
\end{equation}
where $\bar{w}_i$ is a vector of the weights related to the $i$th \gls{acr:pc} at measured $\bar{s}^*_\text{IVC}$ in the training set and $K_{\bar{s}^*_\text{IVC}} := K( \bar{s}^*_\text{IVC},\, \bar{s}^*_\text{IVC},\,\phi ) + \varphi_\text{n} I$.

At last, the scaled expected value and scaled covariance matrix are descaled to complete the description of (\ref{eq:gpr_w}). The descaled expect value is given by
\begin{equation}
    \hat{w}_i(\bar{s}_\text{ICC}) = \hat{\bar{w}}_i(\bar{s}_\text{ICC}) \bar{\sigma}_{w_i} +  \bar{\mu}_{w_i}
\end{equation}
and the descaled covariance matrix is given by
\begin{equation}
    W_{ii}(\bar{s}_\text{ICC}) = \bar{W}_{ii}(\bar{s}_\text{ICC}) \bar{\sigma}_{w_i}.
\end{equation}

\subsection{Reconstructing the In-Cylinder Pressure with Cycle-to-Cycle Variation}
The \glspl{acr:pc} $f(\theta)$ (Section~\ref{sec:pca}) and the estimate behaviour of $w(s_\text{ICC})$ (Section~\ref{sec:gpr}) can be combined to reconstruct a predicted in-cylinder pressure $p(\theta,\, s_\text{ICC})$. Using (\ref{eq:pcd_pressure}), the mean and variance of the in-cylinder pressure can be described by
\begin{equation} \label{eq:expectedProcess}
    \mathbb{E}\left[ p(\theta,\,s_\text{ICC}) \right] = \hat{w}^\trans(s_\text{ICC}) f(\theta) + f_\text{mot}(\theta,\,s_\text{ICC})
\end{equation}
and
\begin{equation} \label{eq:expectedVariance}
    \mathbb{E}\left[ \left( p(\theta,\,s_\text{ICC}) - \mathbb{E}[ p(\theta,\,s_\text{ICC}) ]\right)^2  \right] = f^\trans(\theta) W(s_\text{ICC}) f(\theta),
\end{equation}
respectively.

\section{Combustion Model Identification} \label{sec:hp_select}
The \gls{acr:pcd} and \gls{acr:gpr} require the selection of the number of \glspl{acr:pc} as well as the kernel type and  hyperparameters. 
The training set is used to determine the \glspl{acr:pc} and values for the hyperparameters, while the validation set is used to determine the required amount of \glspl{acr:pc} $n_\text{PC}$ and the best performing kernel type. For this selection, an assessment is made on the prediction accuracy of combustion measures that are relevant for control. To this end,
the \gls{acr:mae} is analysed, which is defined as
\begin{equation}
    \text{MAE}(z) := \frac{1}{n_\text{val} n_\text{cyc}} \sum_{k = 1}^{n_\text{val} n_\text{cyc}} \left| z_{k,\text{meas}} - z_{k,\text{model}} \right|,
\end{equation}
where $n_\text{val}$ is the number of validation measurements, and $z_{\text{meas}}$ and $z_{\text{model}}$ are the combustion metrics resulting from the measured in-cylinder pressure or modelled in-cylinder pressure, respectively. The following combustion measures are studied:
\begin{itemize}
    \item gross Indicated Mean Effective Pressure
        \begin{equation}
            \text{IMEP}_\text{g} = \frac{1}{V_\text{d}} \int_{\theta=\SI{-180}{\degree}}^{\theta=\SI{180}{\degree}} p(\theta)\,dV(\theta)
        \end{equation}
        with displacement volume $V_\text{d}$;
    \item peak pressure $\max(p(\theta))$;
    \item peak pressure rise-rate $\max\left(\tfrac{dp}{d\theta}\right)$;
    \item crank angle where 50\% of the total heat is released 
        \begin{equation}
            \text{CA50} = \left\{\theta \,\left|\, \frac{Q(\theta)}{\max\left(Q(\theta)\right)} = 0.5\right.\right\}
        \end{equation}
        with the heat release \cite{wilhelmsson_model_2006}
        \begin{equation}
            Q(\theta) = \frac{1}{\kappa - 1}p(\theta) V(\theta) + \int_{\alpha=\SI{-180}{\degree}}^{\alpha=\theta} p(\alpha) \frac{dV}{d\alpha} d\alpha - \frac{1}{\kappa - 1}p(\SI{-180}{\degree})V(\SI{-180}{\degree});
        \end{equation} 
    \item burn duration $\text{CA75} - \text{CA25}$ with CA75 and CA25 compute in a similar fashion as CA50;
    \item and burn ratio \cite{willems_heat_2020}
        \begin{equation}
            R_\text{b} = \frac{\text{CA75} - \text{CA50}}{\text{CA50} - \text{CA10}}.
        \end{equation}
\end{itemize}

\begin{figure}
    \centering
    \begin{subfigure}[b]{0.49\textwidth}
        \includegraphics{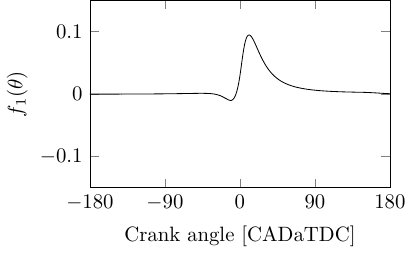}
        \caption{Most relevant \acrshort{acr:pc}}
    \end{subfigure}
    \hfill
    \begin{subfigure}[b]{0.49\textwidth}
        \includegraphics{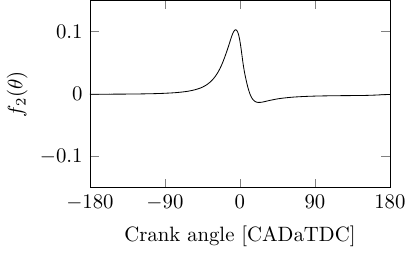}
        \caption{Second most relevant \acrshort{acr:pc}}
    \end{subfigure}
    
    \begin{subfigure}[b]{0.49\textwidth}
        \vspace{1em}
        \includegraphics{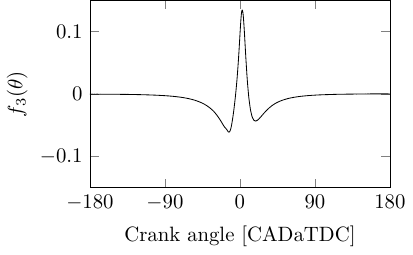}
        \caption{Third most relevant \acrshort{acr:pc}}
    \end{subfigure}
    \hfill
    \begin{subfigure}[b]{0.49\textwidth}
        \vspace{1em}
        \includegraphics{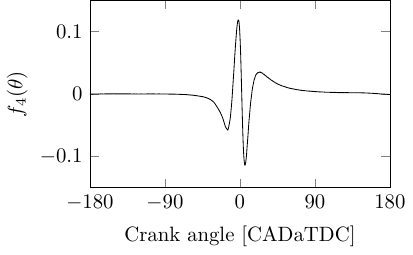}
        \caption{Fourth most relevant \acrshort{acr:pc}}
    \end{subfigure}
    \glsreset{acr:pc}
    \caption{Four most relevant \glspl{acr:pc} resulting from the used training data} \label{fig:pcs}
\end{figure}
\begin{figure}
    \centering
    \begin{subfigure}[b]{0.49\textwidth}
        \includegraphics{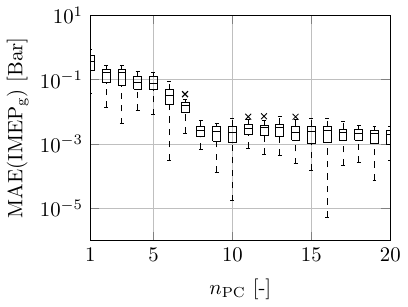}
        \caption{Gross indicated mean effective pressure}
    \end{subfigure}
    \hfill
    \begin{subfigure}[b]{0.49\textwidth}
        \includegraphics{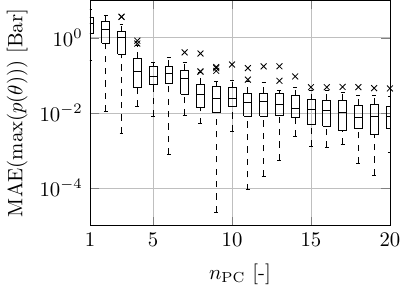}
        \caption{In-cylinder peak pressure}
    \end{subfigure}
    
    \begin{subfigure}[b]{0.49\textwidth}
        \vspace{1em}
        \includegraphics{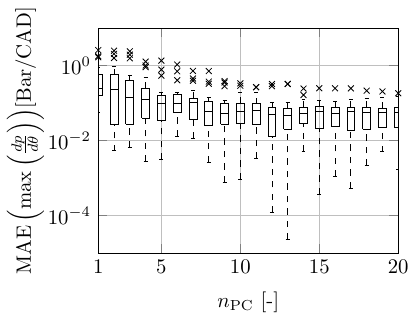}
        \caption{In-cylinder peak pressure rise rate}
    \end{subfigure}
    \hfill
    \begin{subfigure}[b]{0.49\textwidth}
        \vspace{1em}
        \includegraphics{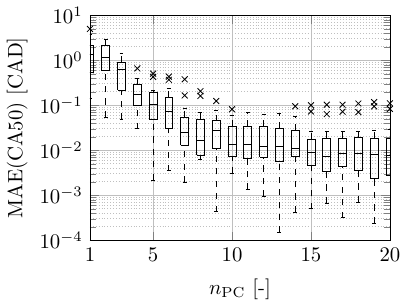}
        \caption{Crank angle at 50\% total heat release}
    \end{subfigure}
    
    \begin{subfigure}[b]{0.49\textwidth}
        \vspace{1em}
        \includegraphics{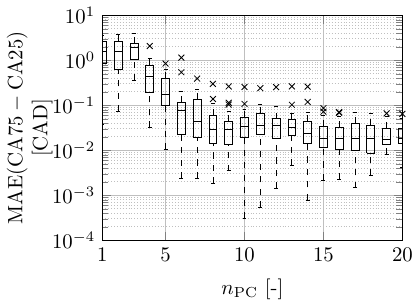}
        \caption{Combustion duration}
    \end{subfigure}
    \hfill
    \begin{subfigure}[b]{0.49\textwidth}
        \vspace{1em}
        \includegraphics{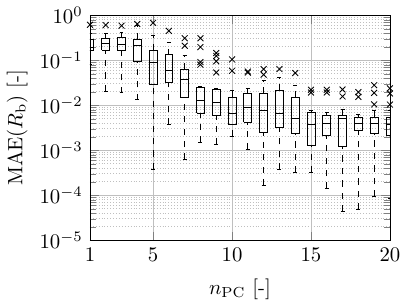}
        \caption{Burn ratio}
    \end{subfigure}
    \caption{Prediction error of combustion metrics for the validation data set using different numbers of Principle Components $n_\text{PC}$. The box plot shows the minimum, maximum, median, first and third quartile, while the crosses show outliers.} \label{fig:nPC}
\end{figure}

\subsection{Selection of Principal Components}
The first hyperparameter is the number of \glspl{acr:pc} $n_\text{PC}$. The \gls{acr:gpr} formulation proposed in Section~\ref{sec:gpr} is not used in this part of the discussion. Figure~\ref{fig:pcs} shows the four most relevant \glspl{acr:pc} derived from the training data, as discussed in Section~\ref{sec:pca}. This figure illustrates that adding more \glspl{acr:pc} will add more higher frequency components to the in-cylinder pressure. Figure~\ref{fig:nPC} shows the absolute error in the corresponding combustion metrics by comparing measurements and model results.
The modelled, decomposed in-cylinder pressure is based on an increasing number of \glspl{acr:pc} using (\ref{eq:pcd_weight}) to compute the required weights. Each measured cycle in the validation set is analysed separately. The figure indicates the minimum, maximum, median, first and third quartile, while the crosses show outliers. It can be seen that the largest gain in improvement is made at lower numbers of \glspl{acr:pc}. From the used training and validation sets, it is concluded that having more than eight \glspl{acr:pc} gives a negligible improvement. Therefore, $n_{PC} = 8$ is used in this study.

\subsection{Selection of Kernel}
Another important aspect in the quality of the model lies in the chosen kernel which describes the correlation between all measured $w(s_\text{ICC}^*)$ and to be predicted mean $\hat{w}(s_\text{ICC})$ and variance $W(s_\text{ICC})$. The kernel types compared in this study rely on the distance measure 
\begin{equation*}
    r(\bar{s}_\text{ICC},\,\bar{s}_\text{ICC}') := \sqrt{ (\bar{s}_\text{ICC} - \bar{s}_\text{ICC}')^\trans \Phi_\text{l}^{-2} (\bar{s}_\text{ICC} - \bar{s}_\text{ICC}') },
\end{equation*}
where $\bar{s}_\text{ICC}$ and $\bar{s}_\text{ICC}'$ are scaled in-cylinder conditions. Each element of the kernel is computed individually. The elements of the kernels used in this work are:
\begin{itemize}
    \item \gls{acr:se}:
        \begin{equation}
            k_\text{SE}(\bar{s}_\text{ICC},\,\bar{s}_\text{ICC}') := \varphi_\text{f}^2 \exp\left( \tfrac{1}{2} r(\bar{s}_\text{ICC},\,\bar{s}_\text{ICC}')^2 \right)
        \end{equation}
        with the set of hyperparameters $\phi = \{\varphi_\text{f},\,\Phi_\text{l}\}$;
    \item Mat\'{e}rn with $\nu = \tfrac{3}{2}$:
        \begin{equation}
            k_\text{Mat\'{e}rn}(\bar{s}_\text{ICC},\,\bar{s}_\text{ICC}') := \varphi_\text{f}^2 \left( 1 + \sqrt{3} r(\bar{s}_\text{ICC},\,\bar{s}_\text{ICC}') \right) \exp\left( -\sqrt{3} r(\bar{s}_\text{ICC},\,\bar{s}_\text{ICC}') \right)
        \end{equation}
        with the set of hyperparameters $\phi = \{\varphi_\text{f},\,\Phi_\text{l}\}$;
    \item Mat\'{e}rn with $\nu = \tfrac{5}{2}$:
        \begin{equation}
            \begin{aligned}
                k_\text{Mat\'{e}rn}(\bar{s}_\text{ICC},\,\bar{s}_\text{ICC}') :=\ &\varphi_\text{f}^2 \left( 1 + \sqrt{5} r(\bar{s}_\text{ICC},\,\bar{s}_\text{ICC}') + 5 r(\bar{s}_\text{ICC},\,\bar{s}_\text{ICC}')^2 \right) \times \\
                &\exp\left( -\sqrt{5} r(\bar{s}_\text{ICC},\,\bar{s}_\text{ICC}') \right)
            \end{aligned}
        \end{equation}
        with the set of hyperparameters $\phi = \{\varphi_\text{f},\,\Phi_\text{l}\}$;
    \item \gls{acr:rq}:
        \begin{equation}
            k_\text{RQ}(\bar{s}_\text{ICC},\,\bar{s}_\text{ICC}') := \varphi_\text{f} \left( \tfrac{1}{2\varphi_\alpha}r(\bar{s}_\text{ICC},\,\bar{s}_\text{ICC}')^2 \right)^{\varphi_\alpha}
        \end{equation}
        with the set of hyperparameters $\phi = \{\varphi_\text{f},\,\varphi_\alpha,\,\Phi_\text{l}\}$.
\end{itemize}
For each kernel, a distinction is made between with and without \gls{acr:ard}. In the case where \gls{acr:ard} is not used, the hyper-parameter $\Phi_\text{l}$ reduces to a scalar. in the case where \gls{acr:ard} is used, the hyper-parameter $\Phi_\text{l}$ is a diagonal matrix with unique elements on the diagonal. The hyper-parameters are determined by maximising the marginal log-likelihood as described in (\ref{eq:mllo}) using the training set.

\begin{table}
    \caption{Mean absolute error in the mean behaviour of important combustion metrics for the validation set using different kernels with $n_\text{PC} = 8$. The best result for each combustion metric is highlighted.} \label{tab:abs_meanvalue}
    \centering
    \scalebox{0.9}{\begin{tabular}{l c c c c c c c c}
        \toprule
        & \multicolumn{4}{c}{without ARD} & \multicolumn{4}{c}{with ARD} \\
        \cmidrule(lr){2-5} \cmidrule(lr){6-9}
        & SE & Mat\'{e}rn & Mat\'{e}rn & RQ & SE & Mat\'{e}rn & Mat\'{e}rn & RQ  \\ 
        &  & $\nu = \tfrac{3}{2}$ & $\nu = \tfrac{5}{2}$ & & & $\nu = \tfrac{3}{2}$ & $\nu = \tfrac{5}{2}$ &\\\midrule
        $\text{IMEP}_\text{g}$ [Bar]                 & 0.2255 & \textbf{0.2061} & 0.2088 & 0.2330 & 0.4161 & 0.2489 & 0.3006 & 0.2769 \\
        $\max(p(\theta))$ [Bar]                      & 2.4564 & \textbf{1.6567} & 1.8007 & 1.9383 & 2.5273 & 1.6653 & 2.0811 & 2.1632 \\ 
        $\max\left(\tfrac{dp}{d\theta}\right)$ [Bar/CAD] & 0.8269 & 0.7880 & 0.7962 & 0.7896 & 0.7546 & \textbf{0.7515} & 0.7987 & 0.7724 \\
        CA50 [CAD]                                   & 0.6121 & 0.5499 & 0.5591 & 0.5489 & 0.9507 & 0.5580 & \textbf{0.5258} & 0.5323 \\
        CA75 - CA25 [CAD]                            & 0.7178 & 0.6795 & 0.6884 & 0.6570 & 0.9371 & 0.5712 & 0.6364 & \textbf{0.5554} \\
        $R_\text{b}$ [-]                             & 0.1456 & 0.1407 & 0.1403 & 0.1325 & 0.2616 & 0.1239 & 0.1669 & \textbf{0.1284} \\
        \bottomrule
    \end{tabular}}
    \vspace{1em}
    \caption{Mean absolute error in the standard deviation of important combustion metrics for the validation set using different kernels with $n_\text{PC} = 8$. The best result for each combustion metric is highlighted.} \label{tab:abs_std}
    \centering
    \scalebox{0.9}{\begin{tabular}{l c c c c c c c c}
        \toprule
        & \multicolumn{4}{c}{without ARD} & \multicolumn{4}{c}{with ARD} \\
        \cmidrule(lr){2-5} \cmidrule(lr){6-9}
        & SE & Mat\'{e}rn & Mat\'{e}rn & RQ & SE & Mat\'{e}rn & Mat\'{e}rn & RQ  \\ 
        &  & $\nu = \tfrac{3}{2}$ & $\nu = \tfrac{5}{2}$ & & & $\nu = \tfrac{3}{2}$ & $\nu = \tfrac{5}{2}$ &\\\midrule
        $\text{IMEP}_\text{g}$ [Bar]                 & 0.4775 & \textbf{0.3280} & 0.3770 & 0.3685 & 0.4960 & 0.4426 & 0.4113 & 0.4210 \\
        $\max(p(\theta))$ [Bar]                      & 1.6716 & \textbf{0.9952} & 1.2239 & 1.1792 & 1.7561 & 1.4948 & 1.3761 & 1.4194 \\
        $\max\left(\tfrac{dp}{d\theta}\right)$ [Bar/CAD] & 0.1177 & 0.1183 & 0.1152 & \textbf{0.1116} & 0.1288 & 0.1461 & 0.1571 & 0.1466 \\
        CA50 [CAD]                                   & 0.2806 & \textbf{0.2261} & 0.2379 & 0.2448 & 0.2664 & 0.2276 & 0.2533 & 0.2329 \\
        CA75 - CA25 [CAD]                            & 0.6130 & 0.5248 & 0.5424 & 0.5327 & 0.5144 & \textbf{0.4510} & 0.4930 & 0.4740 \\
        $R_\text{b}$ [-]                             & 0.1340 & \textbf{0.1296} & 0.1340 & 0.1355 & 0.2616 & 0.1393 & 0.1518 & 0.1584 \\
        \bottomrule
    \end{tabular}}
\end{table}
For the studied combustion measures, Tables~\ref{tab:abs_meanvalue} and \ref{tab:abs_std} show the mean absolute error in the mean behaviour and in the standard deviation, respectively. For each combustion metric, the best result is highlighted. In some cases the difference between the best and second best option are negligible. The Mat\'{e}rn kernel with $\nu = \tfrac{3}{2}$ gives the best result for the most combustion metrics in both mean behaviour and the standard deviation for the used data sets. The resulting \gls{acr:mae} of the mean-value behaviour shows a comparable or improved modelling error as found in literature \cite{klos_investigation_2015,bekdemir_controloriented_2015,khodadadisadabadi_modeling_2016,guardiola_combustion_2018,raut_dynamic_2018,pan_unsupervised_2019,basina_datadriven_2020,kakoee_modeling_2020,xia_robust_2020}.

\section{Validation of the Prediction Quality of the Combustion Model} \label{sec:validation}
\begin{table}
    \centering
    \caption{Selected hyperparameters and kernel used during the validation in Section~\ref{sec:validation}} \label{tab:mdl_hp}
    \begin{tabular}{lc}
        \toprule
        Parameter & Value \\ \midrule
        $n_\text{train}$ & 75 \\
        $n_\text{cyc}$ & 50 \\
        $n_\text{PC}$ & 8 \\
        Kernel & Mat\'{e}rn with $\nu = \tfrac{3}{2}$ \\
        \bottomrule
    \end{tabular}
    \end{table}
\begin{table}
    \caption{Nominal operating conditions of the simulated model for the results shown in Figures~\ref{fig:SOIsweep} and \ref{fig:Timsweep}. For reference, the ranges in experiments are indicated.} \label{tab:soisweep_nominal}
    \centering
    \begin{tabular}{ l c c }
        \toprule
        & Simulated & Measured \\ \midrule
        $Q_\text{tot}$ [kJ] & 2.3 & 2.2 to 2.4 \\
        BR [-] & 0.8 & 0.75 to 0.85 \\
        $\text{SOI}_\text{DI}$ [CADaTDC] & 40 & 40 \\
        $p_\text{im}$ [Bar] & 1.55 & 1.45 to 1.65 \\
        $T_\text{im}$ [\si{\degree C}] & 45 & 40 to 50 \\
        $X_\text{EGR}$ [-] & 0.2 & 0.1 to 0.3
        \\ \bottomrule
    \end{tabular}
\end{table}
The main goal of this work is to predict the in-cylinder pressure and cycle-to-cycle variation. In this section, the outcome of the model is compared to measurements using the validation data set. The hyperparameters shown in Table~\ref{tab:mdl_hp} are used. These choices for hyperparameters give the overall best prediction for the used data set, as discussed in Section~\ref{sec:hp_select}. 

\begin{figure}
    \centering
    \begin{subfigure}[t]{0.49\textwidth}
        \includegraphics{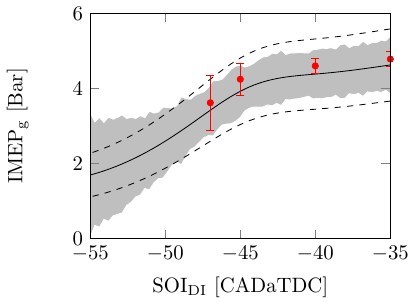}
        \caption{Gross indicated mean effective pressure (solid) and 5\% $\text{cov}\left(\text{IMEP}_\text{g}\right)$ (dashed)}
    \end{subfigure}
    \hfill
    \begin{subfigure}[t]{0.49\textwidth}
        \includegraphics{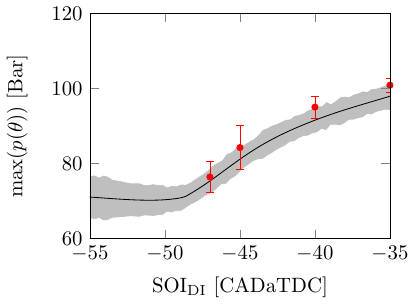}
        \caption{In-cylinder peak pressure}
    \end{subfigure}
    
    \begin{subfigure}[b]{0.49\textwidth}
        \vspace{1em}
        \includegraphics{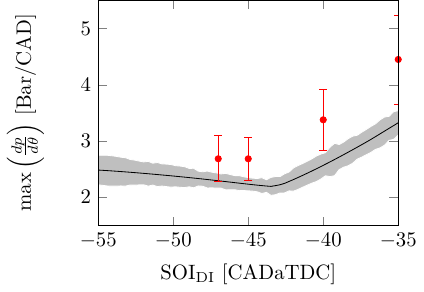}
        \caption{In-cylinder peak pressure rise rate}
    \end{subfigure}
    \hfill
    \begin{subfigure}[b]{0.49\textwidth}
        \vspace{1em}
        \includegraphics{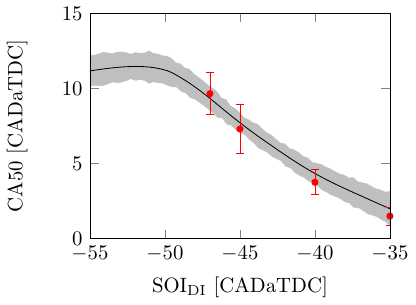}
        \caption{Crank angle at 50\% total heat release}
    \end{subfigure}
    
    \begin{subfigure}[b]{0.49\textwidth}
        \vspace{1em}
        \includegraphics{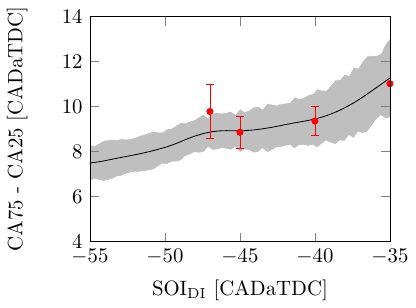}
        \caption{Combustion duration}
    \end{subfigure}
    \hfill
    \begin{subfigure}[b]{0.49\textwidth}
        \vspace{1em}
        \includegraphics{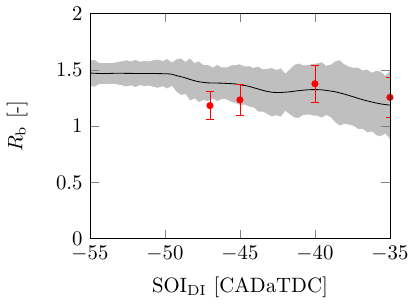}
        \caption{Burn ratio}
    \end{subfigure}
    \caption{Average and cycle-to-cycle variation of important combustion measures (black) and the measured distribution (red) for different values of $\text{SOI}_\text{DI}$ and the nominal conditions shown in Table~\ref{tab:soisweep_nominal} using the hyperparameters as shown in Table~\ref{tab:mdl_hp}.} \label{fig:SOIsweep}
\end{figure}
\begin{table}
    \centering
    \caption{The quality of the predictions for each combustion measure in Figure~\ref{fig:SOIsweep} where four options are distinguished: 1) prediction is correct ($\checkmark$), 2) the trend is followed but the predicted values are too high ($\uparrow$), 3) the trend is followed but predicted values are too low ($\downarrow$), and 4) the prediction is incorrect ($\times$).} \label{tab:SOIsweep}
    \begin{tabular}{l cccccc}
        \toprule
        & $\text{IMEP}_\text{g}$ & $\max(p(\theta))$ & $\max\left(\tfrac{dp}{d\theta}\right)$ & CA50 & $\text{CA75} - \text{CA25}$ & $R_\text{b}$ \\ \midrule
        Mean-value & $\checkmark$ & $\checkmark$ & $\downarrow$ & $\checkmark$ & $\checkmark$ & $\times$ \\ 
        Variance         & $\times$ & $\checkmark$ & $\downarrow$ & $\times$ & $\times$ & $\uparrow$ \\ \bottomrule
    \end{tabular}
\end{table}

\subsection{Variation in Start-of-Injection Directly Injected Fuel}
Figure~\ref{fig:SOIsweep} shows the modelled mean-value and cycle-to-cycle variation of important combustion parameters over a range of $\text{SOI}_\text{DI}$ and the nominal conditions shown in Table~\ref{tab:soisweep_nominal}. The quality of the prediction is classified in Table~\ref{tab:SOIsweep}. Except for the peak-pressure rise rate, the mean-value of the model is similar to that of the measurements. The modelled trend of the peak-pressure rise rate seems to correspond the the measured values. The standard deviation of the model only matches with $\max(p(\theta))$. The trend of the standard deviation of the model of $\max\left(\tfrac{dp}{d\theta}\right)$ and $R_\text{b}$ seems correct, but it is either too high or too low. The standard deviation of the model does not match the measurements for the $\text{IMEP}_\text{g}$, CA50 and $\text{CA75} - \text{CA25}$.  

\begin{figure}
    \centering
    \begin{subfigure}[t]{0.49\textwidth}
        \includegraphics{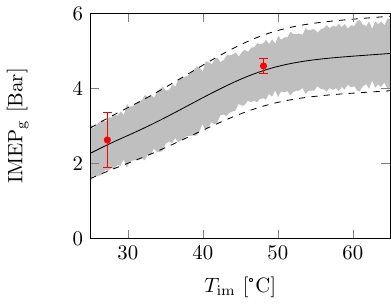}
        \caption{Gross indicated mean effective pressure (solid) and 5\% $\text{cov}(\text{IMEP}_\text{g})$ (dashed)}
    \end{subfigure}
    \hfill
    \begin{subfigure}[t]{0.49\textwidth}
        \includegraphics{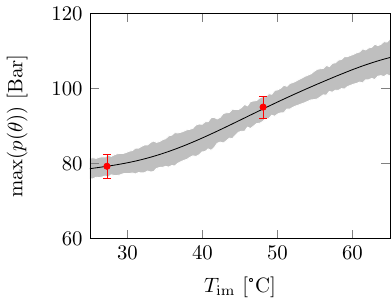}
        \caption{In-cylinder peak pressure}
    \end{subfigure}
    
    \begin{subfigure}[b]{0.49\textwidth}
        \vspace{1em}
        \includegraphics{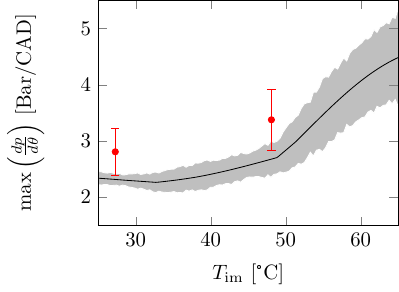}
        \caption{In-cylinder peak pressure rise rate}
    \end{subfigure}
    \hfill
    \begin{subfigure}[b]{0.49\textwidth}
        \vspace{1em}
        \includegraphics{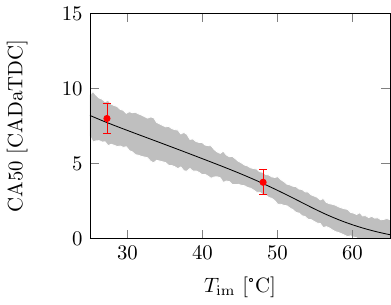}
        \caption{Crank angle at 50\% total heat release}
    \end{subfigure}
    
    \begin{subfigure}[b]{0.49\textwidth}
        \vspace{1em}
        \includegraphics{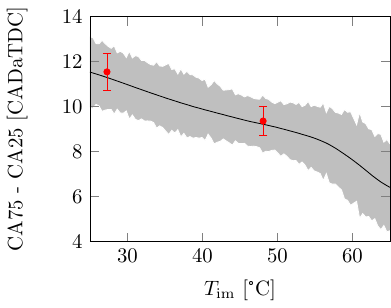}
        \caption{Combustion duration}
    \end{subfigure}
    \hfill
    \begin{subfigure}[b]{0.49\textwidth}
        \vspace{1em}
        \includegraphics{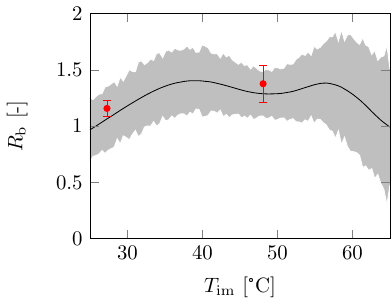}
        \caption{Burn ratio}
    \end{subfigure}
    \caption{Average and cycle-to-cycle variation of important combustion measures (black) and the measured distribution (red) for different values of $T_\text{im}$ and the nominal conditions shown in Table~\ref{tab:soisweep_nominal} using the hyperparameters as shown in Table~\ref{tab:mdl_hp}.} \label{fig:Timsweep}
\end{figure}
\begin{table}
    \centering
    \caption{The quality of the predictions for each combustion measure in Figure~\ref{fig:Timsweep} where four options are distinguished: 1) prediction is correct ($\checkmark$), 2) the trend is followed but the predicted values are too high ($\uparrow$), 3) the trend is followed but predicted values are too low ($\downarrow$), and 4) the prediction is incorrect ($\times$).} \label{tab:Timsweep}
    \begin{tabular}{l cccccc}
        \toprule
        & $\text{IMEP}_\text{g}$ & $\max(p(\theta))$ & $\max\left(\tfrac{dp}{d\theta}\right)$ & CA50 & $\text{CA75} - \text{CA25}$ & $R_\text{b}$ \\ \midrule
        Mean-value & $\checkmark$ & $\checkmark$ & $\downarrow$& $\checkmark$ & $\checkmark$ & $\checkmark$ \\ 
        Variance         & $\times$ & $\checkmark$ & $\times$ & $\checkmark$ & $\uparrow$ & $\times$ \\ \bottomrule
    \end{tabular}
\end{table}

\subsection{Variation in Intake Manifold Temperature}
Figure~\ref{fig:Timsweep} shows the modelled mean-value and cycle-to-cycle variation of important combustion parameters over a range of $T_\text{im}$ and the nominal conditions shown in Table~\ref{tab:soisweep_nominal}. The quality of the prediction is classified in Table~\ref{tab:Timsweep}. Similarly to the sweep of $\text{SOI}_\text{DI}$, the mean-value of the model is similar to that of the measurements except for the peak-pressure rise rate. The modelled trend of the peak-pressure rise rate seems to correspond the the measured values. The standard deviation of the model only matches with $\max(p(\theta))$ and CA50. The trend of the standard deviation of the model of $\text{CA75} - \text{CA25}$ seems correct, but it is too high. The standard deviation of the model does not match the measurements for the $\text{IMEP}_\text{g}$, $\max\left(\tfrac{dp}{d\theta}\right)$ and $R_\text{b}$.

\subsection{Discussion}
In both sweeps, the predicted standard deviations do not always match the measurements. In (\ref{eq:gpr_w}), $w_i(s_{\text{IVC}})$ and $w_j(s_{\text{IVC}})$ $\forall i,\,j \in \{1,\,2,\,\dots,\,\tilde{m}\}$ are assumed to be independent to align with the available \gls{acr:gpr} literature; however, this independence is not necessarily the case. To evaluate the correlation between weights at a fixed $s_\text{ICC}$, the Pearson correlation matrix $R$ is used. This is given by:
\begin{equation}
    [R(s_\text{ICC})]_{ab} = \frac{ \sum_{k=1}^{n_\text{cyc}} \left(  w_{a,k}(s_\text{ICC}) - \tilde{\mu}_{w_a}(s_\text{ICC}) \right) \left(  w_{b,k}(s_\text{ICC}) - \tilde{\mu}_{w_b}(s_\text{ICC}) \right) }{ \tilde{\sigma}_{w_a}(s_\text{ICC})  \tilde{\sigma}_{w_b}(s_\text{ICC}) },
\end{equation}
where $\tilde{\mu}_{w_i}(s_\text{ICC})$ and $\tilde{\sigma}_{w_i}(s_\text{ICC})$ are the mean and standard deviation of the measured weights at $s_\text{ICC}$, respectively. The values of $R$ range from -1 to 1. When an element of R is zero, there is no correlation between the two variables. However, when an element is -1 or 1 there is full correlation between the two variables. The determinant of the $R$ can be used as a measure for the amount of correlation, where $\det(R)$ ranges from 0 to 1. If $\det(R) = 1$ all variables are fully uncorrelated. However, if $\det(R) = 0$ at least two variables are fully correlated.

\begin{figure}
    \centering
    \includegraphics{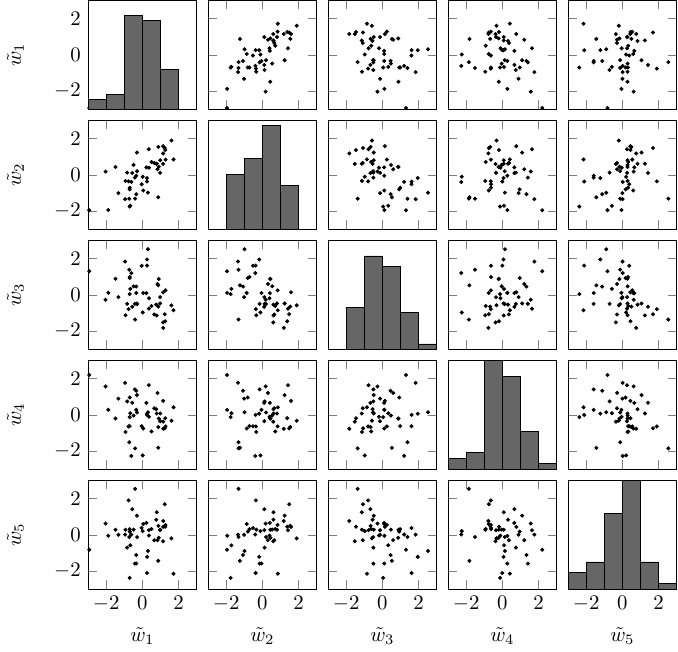}
    \caption{Distribution of the weights for $n_\text{cyc} = 50$ cycles for the first five \glspl{acr:pc} for a constant $s^*_\text{ICC} \in \mathcal{S}^*$ with the least amount of coupling according to the Pearson correlation matrix.} \label{fig:weight_dist}
\end{figure}
Figure~\ref{fig:weight_dist} shows the distribution of the weights for 50 consecutive cycles running at a constant $s^*_\text{ICC} \in \mathcal{S}^*$ with the least amount of coupling according to the determinant of the Pearson correlation matrix. In Figure~\ref{fig:weight_dist}, the weights have been scaled as
\begin{equation*}
    \tilde{w}_ i(s_\text{ICC}) = \frac{ w_i(s_\text{ICC})  - \tilde{\mu}_{w_i}(s_\text{ICC})  }{\tilde{\sigma}_{w_i}(s_\text{ICC})}
\end{equation*}
to emphasise the coupling. The corresponding symmetric Pearson correlation matrix is given by
\begin{equation*}
    R = 
    \begin{bmatrix}
         1      &  0.6762 & -0.3265 & -0.2830 &  0.0240 \\
                &  1      & -0.5037 &  0.0088 &  0.2896 \\
                &         &  1      &  0.1368 & -0.3906 \\
                &         &         &  1      & -0.2157 \\
                &         &         &         &  1      
    \end{bmatrix}
\end{equation*}
with $\det(R) = 0.23$. This shows that the distribution between some of the weights are significantly correlated, as is also illustrated in Figure~\ref{fig:weight_dist}. Therefore, it is no surprise that the quality of the prediction of the cycle-to-cycle variation deviates in the proposed model. This emphasises the importance of developing \gls{acr:gpr} methods that include correlation between the outputs.

\section{Conclusions}
In this study, a data-based model for the in-cylinder pressure and corresponding cycle-to-cycle variations is proposed. This model combines a \gls{acr:pcd} of the in-cylinder pressure and \gls{acr:gpr} to map in-cylinder conditions to a resulting in-cylinder pressure and the corresponding size of the cycle-to-cycle variation. 

In the presented approach, the correlation between $w_i(s_{\text{IVC}})$ and $w_j(s_{\text{IVC}})$ has been neglected for ease of implementation. To improve the accuracy of the cycle-to-cycle variations this correlation should be added. However, there are very few approaches that extend the \gls{acr:gpr} framework to including correlation between model outputs known in literature. 

The proposed data-based modelling approach is successfully applied to an experimental \gls{acr:rcci} engine set-up. The assumption that the model can be split in a general principal component part and operating condition dependent weights is confirmed. A detailed analysis of the hyperparameters for the \gls{acr:pcd} and \gls{acr:gpr} has been performed. It was found that for the used data set more than eight \glspl{acr:pc} do not improve the accuracy of the decomposition based on important combustion measures. For the \gls{acr:gpr}, the Mat\'{e}rn kernel with $\nu = \tfrac{3}{2}$ and without \gls{acr:ard} gives the best results.
The prediction quality of the evaluated combustion measures has an overall accuracy of 13.5\% and 65.5\% in mean behaviour and standard deviation, respectively. The peak-pressure rise-rate is traditionally hard to predict, in the proposed model it has an accuracy of 22.7\% and 96.4\% in mean behaviour and standard deviation, respectively.

In conclusion, the mean-value performance of our model is comparable or shows improvements compared to models found in literature. This shows that, even when neglecting correlation, the model performs well. The model can be used for in-cylinder pressure shaping as proposed in \citeauthor{vlaswinkel_cylinder_2023} \cite{vlaswinkel_cylinder_2023}. Furthermore, it can be used in model-based optimisation approaches that take into account cycle-to-cycle variations and safety criteria. When combined with the \gls{acr:pcd}-based emission model of \citeauthor{henningsson_virtual_2012} \cite{henningsson_virtual_2012}, the model provides a base for optimisation approaches with emission constraints.

\section*{Acknowledgements}
The research presented in this study is financially supported by the Dutch Technology Foundation (STW) under project number 14927. The authors would like to thank Marnix Hage, Michel Cuijpers and Bart van Pinxten from the Zero Emission Lab at the Eindhoven University of Technology for their support during experimentation. The authors would like to thank George Kour for making his \LaTeX-template available at \url{https://github.com/kourgeorge/arxiv-style}.



\vspace{6pt} 

\bibliographystyle{IEEEtranN}
\bibliography{months,library}

\begin{thebibliography}{25}
\providecommand{\natexlab}[1]{#1}
\providecommand{\url}[1]{#1}
\csname url@samestyle\endcsname
\providecommand{\newblock}{\relax}
\providecommand{\bibinfo}[2]{#2}
\providecommand{\BIBentrySTDinterwordspacing}{\spaceskip=0pt\relax}
\providecommand{\BIBentryALTinterwordstretchfactor}{4}
\providecommand{\BIBentryALTinterwordspacing}{\spaceskip=\fontdimen2\font plus
\BIBentryALTinterwordstretchfactor\fontdimen3\font minus \fontdimen4\font\relax}
\providecommand{\BIBforeignlanguage}[2]{{%
\expandafter\ifx\csname l@#1\endcsname\relax
\typeout{** WARNING: IEEEtranN.bst: No hyphenation pattern has been}%
\typeout{** loaded for the language `#1'. Using the pattern for}%
\typeout{** the default language instead.}%
\else
\language=\csname l@#1\endcsname
\fi
#2}}
\providecommand{\BIBdecl}{\relax}
\BIBdecl

\bibitem[Leach et~al.(2020)Leach, Kalghatgi, Stone, and Miles]{leach_scope_2020a}
F.~Leach, G.~Kalghatgi, R.~Stone, and P.~Miles, ``\BIBforeignlanguage{en}{The scope for improving the efficiency and environmental impact of internal combustion engines},'' \emph{\BIBforeignlanguage{en}{Transportation Engineering}}, vol.~1, p. 100005, June 2020.

\bibitem[Duarte Souza Alvarenga~Santos et~al.(2021)Duarte Souza Alvarenga~Santos, R{\"u}ckert~Roso, Teixeira~Malaquias, and Coelho~Ba{\^e}ta]{duartesouzaalvarengasantos_internal_2021}
N.~Duarte Souza Alvarenga~Santos, V.~R{\"u}ckert~Roso, A.~C. Teixeira~Malaquias, and J.~G. Coelho~Ba{\^e}ta, ``\BIBforeignlanguage{en}{Internal combustion engines and biofuels: {{Examining}} why this robust combination should not be ignored for future sustainable transportation},'' \emph{\BIBforeignlanguage{en}{Renewable and Sustainable Energy Reviews}}, vol. 148, p. 111292, September 2021.

\bibitem[Benajes et~al.(2024)Benajes, Garc{\'i}a, {Monsalve-Serrano}, and {Guzm{\'a}n-Mendoza}]{benajes_review_2024}
J.~Benajes, A.~Garc{\'i}a, J.~{Monsalve-Serrano}, and M.~{Guzm{\'a}n-Mendoza}, ``\BIBforeignlanguage{en}{A review on low carbon fuels for road vehicles: {{The}} good, the bad and the energy potential for the transport sector},'' \emph{\BIBforeignlanguage{en}{Fuel}}, vol. 361, p. 130647, April 2024.

\bibitem[Dempsey et~al.(2014)Dempsey, Walker, Gingrich, and Reitz]{dempsey_comparison_2014}
A.~B. Dempsey, N.~R. Walker, E.~Gingrich, and R.~D. Reitz, ``Comparison of {{Low Temperature Combustion Strategies}} for {{Advanced Compression Ignition Engines}} with a {{Focus}} on {{Controllability}},'' \emph{Combustion Science and Technology}, vol. 186, no.~2, pp. 210--241, Februari 2014.

\bibitem[Reitz and Duraisamy(2015)]{reitz_review_2015}
R.~D. Reitz and G.~Duraisamy, ``\BIBforeignlanguage{en}{Review of high efficiency and clean reactivity controlled compression ignition ({{RCCI}}) combustion in internal combustion engines},'' \emph{\BIBforeignlanguage{en}{Progress in Energy and Combustion Science}}, vol.~46, pp. 12--71, Februari 2015.

\bibitem[Paykani et~al.(2021)Paykani, Garcia, Shahbakhti, Rahnama, and Reitz]{paykani_reactivity_2021}
A.~Paykani, A.~Garcia, M.~Shahbakhti, P.~Rahnama, and R.~D. Reitz, ``\BIBforeignlanguage{en}{Reactivity controlled compression ignition engine: {{Pathways}} towards commercial viability},'' \emph{\BIBforeignlanguage{en}{Applied Energy}}, vol. 282, p. 116174, Januari 2021.

\bibitem[Willems(2018)]{willems_cylinder_2018}
F.~Willems, ``\BIBforeignlanguage{en}{Is {{Cylinder Pressure-Based Control Required}} to {{Meet Future HD Legislation}}?}'' \emph{\BIBforeignlanguage{en}{IFAC-PapersOnLine}}, vol.~51, no.~31, pp. 111--118, Januari 2018.

\bibitem[Verhaegh et~al.(2022)Verhaegh, Kupper, and Willems]{verhaegh_datadriven_2022}
J.~Verhaegh, F.~Kupper, and F.~Willems, ``\BIBforeignlanguage{en}{Data-{{Driven Air-Fuel Path Control Design}} for {{Robust RCCI Engine Operation}}},'' \emph{\BIBforeignlanguage{en}{Energies}}, vol.~15, no.~6, p. 2018, Januari 2022.

\bibitem[Khodadadi~Sadabadi et~al.(2016)Khodadadi~Sadabadi, Shahbakhti, Bharath, and Reitz]{khodadadisadabadi_modeling_2016}
K.~Khodadadi~Sadabadi, M.~Shahbakhti, A.~N. Bharath, and R.~D. Reitz, ``\BIBforeignlanguage{en}{Modeling of combustion phasing of a reactivity-controlled compression ignition engine for control applications},'' \emph{\BIBforeignlanguage{en}{International Journal of Engine Research}}, vol.~17, no.~4, pp. 421--435, April 2016.

\bibitem[Guardiola et~al.(2018)Guardiola, Pla, Bares, and Barbier]{guardiola_combustion_2018}
C.~Guardiola, B.~Pla, P.~Bares, and A.~Barbier, ``\BIBforeignlanguage{en}{A combustion phasing control-oriented model applied to an {{RCCI}} engine},'' \emph{\BIBforeignlanguage{en}{IFAC-PapersOnLine}}, vol.~51, no.~31, pp. 119--124, Januari 2018.

\bibitem[Raut et~al.(2018)Raut, Irdmousa, and Shahbakhti]{raut_dynamic_2018}
A.~Raut, B.~K. Irdmousa, and M.~Shahbakhti, ``\BIBforeignlanguage{en}{Dynamic modeling and model predictive control of an {{RCCI}} engine},'' \emph{\BIBforeignlanguage{en}{Control Engineering Practice}}, vol.~81, pp. 129--144, December 2018.

\bibitem[Kakoee et~al.(2020)Kakoee, Bakhshan, Barbier, Bares, and Guardiola]{kakoee_modeling_2020}
A.~Kakoee, Y.~Bakhshan, A.~Barbier, P.~Bares, and C.~Guardiola, ``\BIBforeignlanguage{en}{Modeling combustion timing in an {{RCCI}} engine by means of a control oriented model},'' \emph{\BIBforeignlanguage{en}{Control Engineering Practice}}, vol.~97, p. 104321, April 2020.

\bibitem[Bekdemir et~al.(2015)Bekdemir, Baert, Willems, and Somers]{bekdemir_controloriented_2015}
C.~Bekdemir, R.~Baert, F.~Willems, and B.~Somers, ``Towards {{Control-Oriented Modeling}} of {{Natural Gas-Diesel RCCI Combustion}},'' in \emph{{{SAE}} 2015 {{World Congress}} \& {{Exhibition}}}.\hskip 1em plus 0.5em minus 0.4em\relax {SAE International}, April 2015, pp. 2015--01--1745.

\bibitem[Klos and Kokjohn(2015)]{klos_investigation_2015}
D.~Klos and S.~L. Kokjohn, ``\BIBforeignlanguage{en}{Investigation of the sources of combustion instability in low-temperature combustion engines using response surface models},'' \emph{\BIBforeignlanguage{en}{International Journal of Engine Research}}, vol.~16, no.~3, pp. 419--440, April 2015.

\bibitem[Xia et~al.(2020)Xia, {de Jager}, Donkers, and Willems]{xia_robust_2020}
L.~Xia, B.~{de Jager}, T.~Donkers, and F.~Willems, ``\BIBforeignlanguage{en}{Robust constrained optimization for {{RCCI}} engines using nested penalized particle swarm},'' \emph{\BIBforeignlanguage{en}{Control Engineering Practice}}, vol.~99, p. 104411, June 2020.

\bibitem[Basina et~al.(2020)Basina, Irdmousa, Velni, Borhan, Naber, and Shahbakhti]{basina_datadriven_2020}
L.~N.~A. Basina, B.~K. Irdmousa, J.~M. Velni, H.~Borhan, J.~D. Naber, and M.~Shahbakhti, ``\BIBforeignlanguage{en}{Data-driven {{Modeling}} and {{Predictive Control}} of {{Maximum Pressure Rise Rate}} in {{RCCI Engines}}},'' in \emph{\BIBforeignlanguage{en}{2020 {{IEEE Conference}} on {{Control Technology}} and {{Applications}} ({{CCTA}})}}.\hskip 1em plus 0.5em minus 0.4em\relax {Montreal, QC, Canada}: {IEEE}, August 2020, pp. 94--99.

\bibitem[Pan et~al.(2019)Pan, Korkmaz, Beeckmann, and Pitsch]{pan_unsupervised_2019}
W.~Pan, M.~Korkmaz, J.~Beeckmann, and H.~Pitsch, ``\BIBforeignlanguage{en}{Unsupervised learning and nonlinear identification for in-cylinder pressure prediction of diesel combustion rate shaping process},'' in \emph{\BIBforeignlanguage{en}{{{IFAC-PapersOnLine}}}}, ser. 13th {{IFAC Workshop}} on {{Adaptive}} and {{Learning Control Systems ALCOS}} 2019, vol.~52, Januari 2019, pp. 199--203.

\bibitem[Vlaswinkel et~al.(2022)Vlaswinkel, De~Jager, and Willems]{vlaswinkel_databased_2022a}
M.~Vlaswinkel, B.~De~Jager, and F.~Willems, ``\BIBforeignlanguage{en}{Data-{{Based In-Cylinder Pressure Model}} including {{Cyclic Variations}} of an {{RCCI Engine}}},'' \emph{\BIBforeignlanguage{en}{IFAC-PapersOnLine}}, vol.~55, no.~24, pp. 13--18, 2022.

\bibitem[Henningsson et~al.(2012)Henningsson, Tunest{\aa}l, and Johansson]{henningsson_virtual_2012}
M.~Henningsson, P.~Tunest{\aa}l, and R.~Johansson, ``\BIBforeignlanguage{en}{A {{Virtual Sensor}} for {{Predicting Diesel Engine Emissions}} from {{Cylinder Pressure Data}}},'' \emph{\BIBforeignlanguage{en}{IFAC Proceedings Volumes}}, vol.~45, no.~30, pp. 424--431, Januari 2012.

\bibitem[Panzani et~al.(2017)Panzani, Ostman, and Onder]{panzani_engine_2017}
G.~Panzani, F.~Ostman, and C.~H. Onder, ``\BIBforeignlanguage{en}{Engine {{Knock Margin Estimation Using In-Cylinder Pressure Measurements}}},'' \emph{\BIBforeignlanguage{en}{IEEE/ASME Transactions on Mechatronics}}, vol.~22, no.~1, pp. 301--311, Februari 2017.

\bibitem[Panzani et~al.(2019)Panzani, Pozzato, Savaresi, R{\"o}sgren, and Onder]{panzani_engine_2019a}
G.~Panzani, G.~Pozzato, S.~M. Savaresi, J.~R{\"o}sgren, and C.~H. Onder, ``\BIBforeignlanguage{en}{Engine knock detection: An eigenpressure approach},'' \emph{\BIBforeignlanguage{en}{IFAC-PapersOnLine}}, vol.~52, no.~5, pp. 267--272, 2019.

\bibitem[Vlaswinkel and Willems(2023)]{vlaswinkel_cylinder_2023}
M.~Vlaswinkel and F.~Willems, ``\BIBforeignlanguage{en}{Cylinder {{Pressure Feedback Control}} for {{Ideal Thermodynamic Cycle Tracking}}: {{Towards Self-learning Engines}}},'' \emph{\BIBforeignlanguage{en}{IFAC-PapersOnLine}}, vol.~56, no.~2, pp. 8260--8265, 2023.

\bibitem[Rasmussen and Williams(2005)]{rasmussen_gaussian_2005}
C.~E. Rasmussen and C.~K.~I. Williams, \emph{Gaussian {{Processes}} for {{Machine Learning}}}.\hskip 1em plus 0.5em minus 0.4em\relax {The MIT Press}, November 2005.

\bibitem[Wilhelmsson et~al.(2006)Wilhelmsson, Tunest, and Johansson]{wilhelmsson_model_2006}
C.~Wilhelmsson, P.~Tunest, and B.~Johansson, ``\BIBforeignlanguage{en}{Model {{Based Engine Control Using ASICs}}: {{A Virtual Heat Release Sensor}}},'' in \emph{\BIBforeignlanguage{en}{Proceedings of the Les Rencontres Scientifiques de l’IFP: "New Trends in Engine Control, Simulation and Modelling"}}.\hskip 1em plus 0.5em minus 0.4em\relax {Institut Francais du Petrole}, 2006.

\bibitem[Willems et~al.(2020)Willems, Willems, Deen, and Somers]{willems_heat_2020}
R.~Willems, F.~Willems, N.~Deen, and B.~Somers, ``\BIBforeignlanguage{en}{Heat release rate shaping for optimal gross indicated efficiency in a heavy-duty {{RCCI}} engine fueled with {{E85}} and diesel},'' \emph{\BIBforeignlanguage{en}{Fuel}}, p. 119656, November 2020.

\end{thebibliography}

\end{document}